\documentclass[prb,twocolumn,showpacs,preprintnumbers,superscriptaddress]{revtex4}

\usepackage{graphicx}
\usepackage{graphics}
\usepackage{amsmath}
\usepackage{amssymb}
\usepackage{color}
\usepackage{dcolumn}
\usepackage{epsfig}
\usepackage{bm}

\begin{document}

\title{Magnetic structure of the edge-sharing copper oxide chain compound NaCu$_{2}$O$_{2}$}

\author{L. Capogna}
\affiliation{Consiglio Nazionale delle Ricerche, IOM-OGG, 6 rue J. Horowitz, F-38042 Grenoble, France}
\affiliation{Institut Laue Langevin, 6 rue J. Horowitz, F-38042 Grenoble, France}
\author{M. Reehuis}
\affiliation{Helmholtz-Zentrum Berlin f\"ur Materialen und Energie, Glienicker Str.100, D-14109 Berlin, Germany}
\author{A. Maljuk}
\affiliation{Helmholtz-Zentrum Berlin f\"ur Materialen und Energie, Glienicker Str.100, D-14109 Berlin, Germany}
\affiliation{Max-Planck-Institut f\"{u}r Festk\"{o}rperforschung, Heisenbergstr. 1, D-70569 Stuttgart, Germany}
\author{R.K. Kremer}
\affiliation{Max-Planck-Institut f\"{u}r Festk\"{o}rperforschung, Heisenbergstr. 1, D-70569 Stuttgart, Germany}
\author{B. Ouladdiaf}
\affiliation{Institut Laue Langevin, 6 rue J. Horowitz, F-38042 Grenoble, France}
\author{M. Jansen}
\affiliation{Max-Planck-Institut f\"{u}r Festk\"{o}rperforschung, Heisenbergstr. 1, D-70569 Stuttgart, Germany}
\author{B. Keimer}
\affiliation{Max-Planck-Institut f\"{u}r Festk\"{o}rperforschung, Heisenbergstr. 1, D-70569 Stuttgart, Germany}

\date{\today}

\begin{abstract}
Single-crystal neutron diffraction has been used to determine the incommensurate magnetic structure of NaCu$_{2}$O$_{2}$, a compound built up of chains of edge-sharing CuO$_4$ plaquettes. Magnetic structures compatible with the lattice symmetry were identified by a group-theoretical analysis, and their magnetic structure factors were compared to the experimentally observed Bragg intensities. In conjunction with other experimental data, this analysis yields an elliptical helix structure in which both the helicity and the polarization plane alternate among copper-oxide chains. This magnetic ground state is discussed in the context of the recently reported multiferroic properties of other copper-oxide chain compounds.
\end{abstract}


\maketitle

\section{Introduction}
Research on compounds with coupled spontaneous magnetic and electric polarization (multiferroics) has recently focused renewed attention on helical magnetism. A leading theory of multiferroicity predicts that helical magnetic order in insulators induces an electric polarization of the form

\begin{equation}
\mathbf{P} \propto \sum_{\langle ij \rangle} \hat{n}_{ij} \times ( \mathbf{S}_{i} \times \mathbf{S}_{j} ),
\end{equation}

\noindent where $\hat{n}_{i,j}$ is a vector connecting spins $\textbf{S}_{i,j}$ on exchange bond $\langle ij \rangle$. \cite{katsura,sergienko,mostovoy} Clearly, detailed information about the magnetic structure and crystal symmetry is required to evaluate the consequences of Eq. 1 for any given compound. The theory explains the observation of ferroelectricity in compounds such as TbMnO$_3$ and CuO, which exhibit helicoidal states with nonvanishing projection of the propagation vector onto the polarization plane. \cite{wang} While these compounds contain three-dimensional (3D) networks of exchange bonds, magnetic insulators with quasi-1D electronic structure have the potential to serve as particularly instructive model systems for multiferroicity. The discovery of a macroscopic electric polarization in the compounds LiCu$_{2}$O$_{2}$ (Ref. \onlinecite{park}) and LiCuVO$_{4}$ (Ref. \onlinecite{naito}), which are built up of chains of edge-sharing CuO$_4$ plaquettes, has therefore generated significant attention. Because of the competition between the anomalously small nearest-neighbor exchange coupling and the stronger next-nearest neighbor coupling of spin-1/2 copper ions along the chains, both compounds exhibit incommensurate magnetic correlations, and interchain interactions induce helicoidal 3D long-range order at low temperatures. In addition to multiferroicity, these materials are also of interest in the context of research on the interplay between spin and charge correlations along the copper-oxide chains. \cite{pisarev,malek,matiks}

In spite of their simple electronic structure, the multiferroic properties of LiCu$_{2}$O$_{2}$ and LiCuVO$_{4}$ are still poorly understood. Both compounds are orthorhombic, with copper-oxide chains running along the $b$-axis. Based on neutron diffraction experiments, the polarization plane of the helix in LiCu$_{2}$O$_{2}$ was originally reported to lie within the $ab$ plane, \cite{masuda} while {\bf P} was found to be along $c$, in disagreement with Eq. 1. \cite{park} Later polarized-neutron experiments \cite{seki} indicated a $bc$-polarized helix, which is consistent with the electrical polarization according to Eq. 1. However, the behavior of {\bf P} in an applied magnetic field {\bf H} was found to be difficult to reconcile with this scenario, \cite{park,yasui1} although interchain magneto-electric coupling may offer a possible solution. \cite{fang} Recent single-crystal neutron diffraction data \cite{kobayashi} have been interpreted as evidence of an elliptical helix state with polarization plane tilted by an angle of 45$^{\circ}$ with respect to the $bc$-plane. The consequences of this structure for the electric polarization remain to be assessed. In LiCuVO$_{4}$, the helix was reported to be polarized within the $ab$-plane for $\mathbf{H} = 0$, \cite{gibson,yasui2} which is consistent with the observation that $\mathbf{P} \parallel a$. However, deviations from the theoretical predictions were again noted for $\mathbf{H} \neq 0$. \cite{yasui2} The apparent failure of Eq. 1 as a description of the magnetic field dependence of the dielectric properties of LiCu$_{2}$O$_{2}$ and LiCuVO$_{4}$ has motivated an alternative scenario according to which the observed electric polarization is due to defects generated by Li--Cu intersubstitution. \cite{moskvin_epl,moskvin_prb} While such defects are indeed quite common due to the similar ionic radii of Li and Cu, \cite{masuda,prokofiev} this scenario has been contested based on work on ostensibly stoichiometric LiCu$_{2}$O$_{2}$ single crystals. \cite{yasui1} The situation thus remains unresolved.

Here we present an investigation of the magnetic structure of NaCu$_{2}$O$_{2}$, a compound that is isostructural and isolelectronic to LiCu$_{2}$O$_{2}$ and, like LiCu$_{2}$O$_{2}$ and LiCuVO$_{4}$, exhibits incommensurate magnetic order at low temperatures. \cite{capogna,horvatic,gippius} Unlike LiCu$_{2}$O$_{2}$ and LiCuVO$_{4}$, however, NaCu$_{2}$O$_{2}$ is not intrinsically prone to disorder because of the different sizes of Na and Cu ions, and Na--Cu intersubstitution can be ruled out with high sensitivity. \cite{capogna} A recent investigation  \cite{leininger} came to the conclusion that ferroelectricity is absent in NaCu$_{2}$O$_{2}$. Detailed information about the magnetic structure is required in order to assess the origin of the qualitatively different electric properties of LiCu$_{2}$O$_{2}$ and LiCuVO$_{4}$ on the one hand, and NaCu$_{2}$O$_{2}$ on the other hand. However, while a comprehensive set of single-crystal neutron diffraction data is available for the former two compounds, \cite{kobayashi,gibson,yasui2} information about the magnetic structure of NaCu$_{2}$O$_{2}$ has thus far only been gleaned from more limited neutron powder diffraction, \cite{capogna} nuclear magnetic resonance (NMR), \cite{horvatic,gippius} and resonant x-ray diffraction \cite{leininger} data. Motivated by the controversy about the origin of multiferrocity in copper-oxide chain compounds outlined above, we have carried out a single-crystal neutron diffraction study of the magnetic structure of NaCu$_{2}$O$_{2}$. We present the results in the framework of a rigorous symmetry classification of possible magnetic structures.

\section{Results and analysis}

Single crystals of NaCu$_{2}$O$_{2}$ in the shape of small rectangular plates a few millimeters long and wide and $\sim 0.1$ mm thick were grown by a self-flux technique described elsewhere. \cite{maljuk} X-ray diffraction, \cite{maljuk} induction-coupled plasma atomic emission spectroscopy, \cite{maljuk} and magnetometry \cite{leininger} were used to fully characterize the structure, purity, and quality of the crystals. Neutron diffraction experiments were carried out at the Institut Laue-Langevin using the high-resolution powder diffractometer D2B, the Laue single-crystal diffractometer Vivaldi, and the four-circle diffractometer D10. For the D2B measurements, we used a powder sample described elesewhere. \cite{capogna}

\begin{figure}[htbp]
\begin{center}
\includegraphics[width=8cm,angle=0]{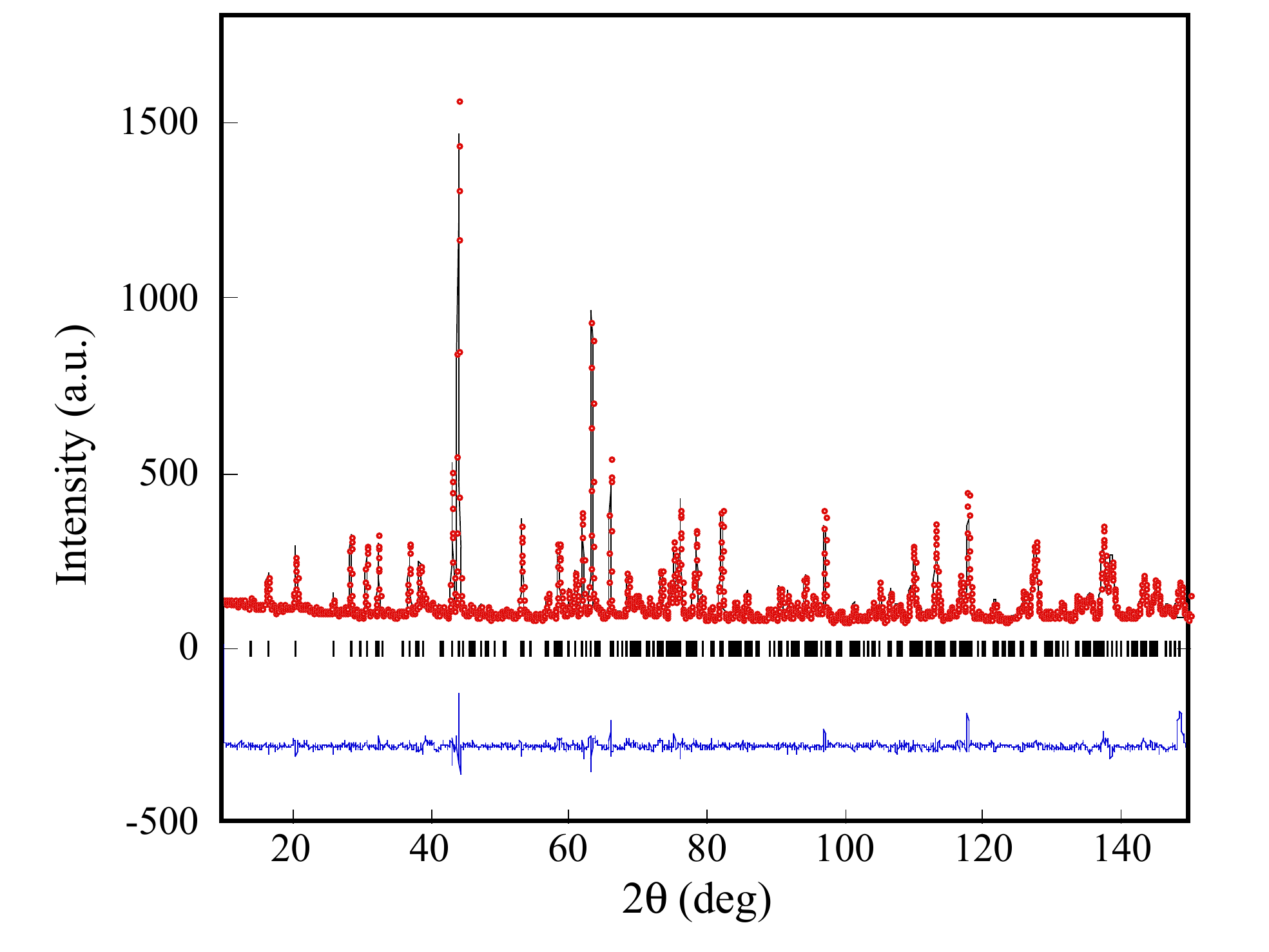}\\
\caption{Powder diffraction data on NaCu$_{2}$O$_{2}$ taken on D2B with a neutron wavelength $\lambda=1.596$ \AA\
at $T=1.5$ K. The line through the data is the result of a standard Rietveld refinement. The lower curve indicates the corresponding residuals. Tick marks indicate the positions of nuclear Bragg reflections.}
\end{center}
\end{figure}

Figure 1 shows powder diffraction data taken on D2B, which were used to accurately refine the nuclear structure at temperatures 1.5 and 300 K. Rietveld refinement in the orthorhombic space group $Pnma$ (No. 62) with parameters shown in Table I yields an excellent description of the data (Fig. 1).

\begin{table}[htbp]
\caption {
Lattice parameters and atomic positions in fractional coordinates, as derived from Rietveld refinement of powder diffraction data at temperatures $T=1.5$ and 300 K. The space group is $Pnma$. All atoms are in Wyckoff position $4c$.}
\centering
\begin{tabular} {l l l l l l}
\hline\hline
\\[0.1mm]
 $T$ (K) & & 1.5   \hspace{5 mm}  & 300 \hspace{5 mm}  \\[0.5mm]
 \hline
 \\ [0.1mm]
 $a$ (\AA) & & 6.2001(1) & 6.2148(1)   \\
 $b$ (\AA)&  & 2.9310(1) &  2.9361(1)  \\
 $c$ (\AA)&  & 13.0337(2) & 13.0731(3)   \\[0.1mm]
 Cu(I)&  $x$ &   0.3810(7) & 0.3811(1) \\
 & $z$ & 0.2549(2) & 0.2556(2) \\ [0.1mm]
 Cu(II)&  $x$   &   0.8694(7)  & 0.8691(8) \\
 & $z$ & 0.6059(2)   & 0.6050(2)\\ [0.1mm]
 Na &  $x$ & 0.358(1) & 0.355(1) \\
 & $z$ & 0.5806(4) & 0.5800(5) \\ [0.1mm]
 O(I) & $x$   &   0.4189(6) & 0.4195(7) \\
 & $z$  & 0.1155(3)  &  0.1160(3) \\ [0.1mm]
 O(II) & $x$  &   0.3489(8) & 0.3491(1) \\
 & $z$ & 0.3979(3)  & 0.3980(3) \\ [0.1mm]
\hline \\ [0.1mm]
$R_F$ & & 0.054 & 0.055
  \\ [0.1mm]
 \hline \hline
\end{tabular}
\label{table:I}
\end{table}

In order to determine the magnetic structure of NaCu$_{2}$O$_{2}$, a set of 40 magnetic reflections was acquired on a single crystal at $T=1.5$ K. The data were taken in a three-axis configuration on D10 with a neutron wavelength of 2.36 \AA. The magnetic reflections were found to be incommensurate with the crystal lattice and could be indexed as $ (h, k, l)_M = (h, k, l)_N \pm {\bf k}$, where $N$ stands for nuclear reflections. The propagation vector ${\bf k} =  (0.5, \xi, 0)$ with $\xi = 0.228 \pm 0.002$ resulting from refinement of the data is consistent with prior neutron powder diffraction \cite{capogna} and resonant x-ray diffraction \cite{leininger} measurements. The temperature dependence of the magnetic Bragg intensity determined on Vivaldi (Fig. 2) showed a N\'{e}el temperature $T_N = 12$ K, again in agreement with prior work. \cite{capogna,leininger} The integrated intensities of the magnetic reflections were determined by rocking scans (inset in Fig. 2). For the refinement of the magnetic structure we used the magnetic structure factors of 20 incommensurate magnetic reflections, 10 of which are unique. The overall scale factor as well as the extinction parameter were fixed to the values refined from a set of nuclear reflections collected under the same experimental conditions. The atomic positions and the lattice parameters were fixed to the values refined from the powder data (Table I).

\begin{figure}[htbp]
\begin{center}
\includegraphics[width=8cm,angle=0]{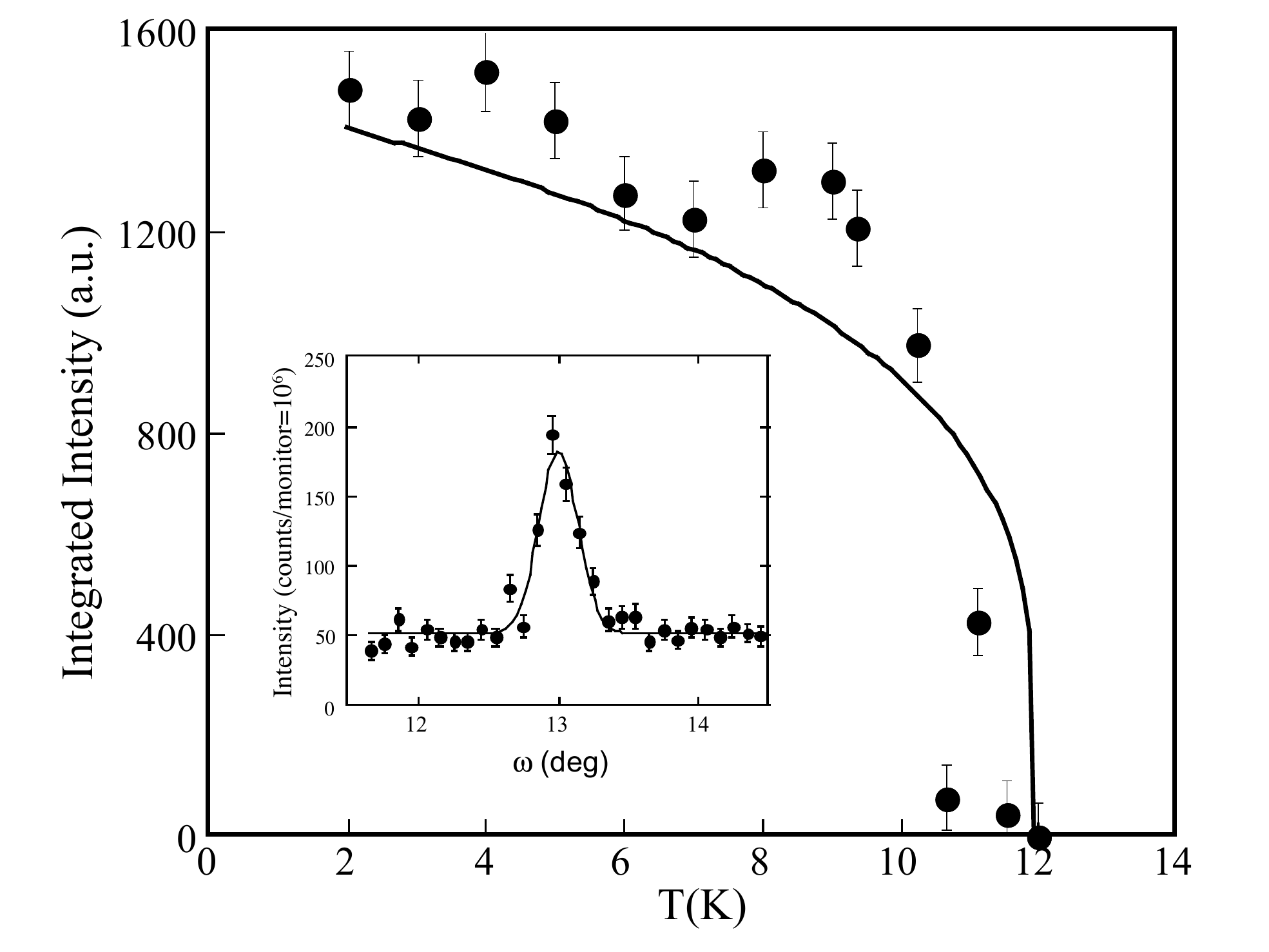}\\
\caption{Temperature dependence of the intensity of the magnetic Bragg reflection $(0.5, -0.228, -1)$ in NaCu$_{2}$O$_{2}$, taken on the Laue diffractometer Vivaldi. The inset shows a rocking curve taken at $T=2$ K on the four-circle diffractometer D10.}
\end{center}
\end{figure}

In order to examine the constraints set on the magnetic structure by the crystal symmetry, we used the representation analysis method \cite{bertaut} as implemented in the program BasIreps. \cite{carvajal} The lattice symmetry (space group $Pnma$) and atomic positions derived from the structural analysis described above, as well as the propagation vector {\bf k} determined from the positions of the magnetic reflections serve as input for this analysis. Further, x-ray absorption measurements indicate valence states of 1+ (with a full-shell configuration), and 2+ (with one unpaired hole in the $d$-shell) for the Cu(I) and Cu(II) ions, respectively. \cite{leininger} Only the latter ions therefore carry a magnetic moment. They are in the following positions:

\begin{flushleft}
Cu1: $(x, \, 1/4, \, z)$ \\
Cu2: $(-x, \, 3/4, \, -z) + (1, \, 0, \, 1)$ \\
Cu3: $(1/2+x, \, 1/4, \, 1/2-z) + (-1, \, 0, \, 1)$ \\
Cu4: $(1/2-x, \, 3/4, \, 1/2+z) + (1, \, 0, \, -1)$
\end{flushleft}

\noindent with refined parameters $x$ and $z$ listed in Table I.

The representation analysis yields a single, two-dimensional irreducible representation, $\Gamma_{mag}$, for the magnetic structure. The possible spin configurations of the four Cu$^{2+}$ ions in the primitive unit cell are described by the basis vectors $\Psi_n$ (with $n= 1 ... 12$) of $\Gamma_{mag}$, which are displayed in Table II. We denote ferro- and antiferromagnetic alignment of spins along the crystallographic axis $i$ by the modes $f_i$ and $a_i$, respectively. The modes $f_x$, $a_y$, and $f_z$ resulting from the representation analysis imply the sign sequences $(+ + + +)$, $(+ - + -)$, and $(+ + - -)$ for the $x$-, $y$-, and $z$-axis components, respectively, of the magnetization of atoms $\rm (Cu1 \; Cu2 \; Cu3 \; Cu4)$. Note that the two-dimensionality of $\Gamma_{mag}$ (which derives from the spin reversal along $a$ implied by the propagation vector) permits a physically equivalent description in terms of the modes $a_x$, $f_y$, and $a_z$ of spins in an adjacent subcell of the magnetic lattice (Table II).

\begin{table}[htbp]
\caption {Basis functions $\Psi_n$ of the irreducible representation $\Gamma_{mag}$
for propagation vector ${\bf k} =  (0.5, 0.228, 0)$ in the orthorhombic space
group $Pnma$. Only the real parts of the basis vectors are given, with $\alpha = \cos 2\pi(k_y/2) = 0.754$.
The modes $f_i$ and $a_i$ characterize ferromagnetic and antiferromagnetic spin configurations, respectively, along crystal axis $i$. }
\centering
\begin{tabular} {c c c c c c c c c}
\hline\hline
\\[0.5mm]
	 & & Cu1 & Cu2 & & Cu3 & Cu4 &\\
	\\[0.01mm]	
\hline
\\[0.01mm]
	&$\Psi_1$ &(100)&($\alpha$00)&$f_x$&(000)&(000)&\\	
	&$\Psi_2$ &(010)&(0$-\alpha$0)&$a_y$&(000)&(000)&\\
	&$\Psi_3$ &(001)&(00$\alpha$)&$f_z$&(000)&(000)&\\
	\\[0.01mm]
	\hline
	\\[0.01mm]	
	&$\Psi_4$ &(000)&(000)	& &(100)&($\alpha$00)&	$f_x$ \\	
	&$\Psi_5$ &(000)&(000)	& &(010)&(0$-\alpha$0)&	$a_y$ \\	
	&$\Psi_6$ &(000)&(000)	& &(00$-1$)&(00$-\alpha$)&	$f_z$ \\
	\\[0.01mm]
	\hline
	\\[0.01mm]	
	&$\Psi_7$ &(000)&(000)	& &($-1$00)&($\alpha$00)&	$a_x$ \\	
	&$\Psi_8$ &(000)&(000)	& &(0$-1$0)&(0$-\alpha$0)&	$f_y$ \\	
	&$\Psi_9$ &(000)&(000)	& &(001)&(00$-\alpha$)&	$a_z$ \\
	\\[0.01mm]
	\hline	
	\\[0.01mm]
	&$\Psi_{10}$ &	(100)&($-\alpha$00)&$a_x$ &(000)&	(000)&	\\	
	&$\Psi_{11}$ &	(010)&(0$\alpha$0)&$f_y$ & (000)&(000)&	\\	
	&$\Psi_{12}$ &	(001)&(00$-\alpha$)&$a_z$ &(000)&	(000)&	\\	
\\
\hline
\hline
 \end{tabular}
 \label{table:II}
 \end{table}

\begin{table*}[htbp]
\caption {Comparison of different models for the magnetic structure of NaCu$_{2}$O$_{2}$.
The imaginary and real components of the Fourier coefficients of the Cu$^{2+}$ magnetic moments, ($I_x$, $I_y$, $I_z$) and ($R_x$, $R_y$, $R_z$), as well as the resulting real part $R$ and total moment $M$ are given in units of $\mu_B$. For the sinusoidal structure, the magnetic moment amplitude is listed ($\ast$). The agreement factors of the least-square fits are defined as $R_F = \sum(|F_{obs}|- |F_{calc}|)/\sum|F_{obs}|$ and $\chi^2 = \sum (|I_{obs} - I_{calc}|/\sigma^2)/(n - p)$, where $n$ and $p$ are the numbers of observations and refined parameters, respectively.}
\begin{tabular}{l c c c c c c c c c c c c c  }
\hline\hline
\\[0.1mm]
Model      &  1 &  & 2 &  & 3 & & 4 & & 5 &  & 6 &  & 7  \\
\\[0.5mm]
\hline
\\[0.5mm]
$I_x$  &      0 &     & 0.48(4) & & 0       &&      0          & & 0 & & 0 & & 0   \\
\\[0.01mm]
$I_y$ &       0 &     & 0       & & 0       & &     0.47(6)    & & 0.49(2) & & 0.54(2)  & & 0  \\
\\[0.01mm]
$I_z$  &      0 &     & 0       & & 0.46(6) & &      0 & &  0  & & 0  & & 0.54(4)  \\
\\[0.01mm]
$R_x$  &      0.39(5) && 0       & & 0.39(5) & &     0.27(7)   & & 0 & & 0.54(2) & & 0.54(4) \\
\\[0.01mm]
$R_y$  &      0.47(5) && 0.46(4) & & 0.48(5) & &      0        & & 0 &  & 0 & & 0  \\
\\[0.01mm]
$R_z$  &      0.46(6) && 0.46(5) & & 0       & &      0.46(8)  & & 0.49(2)	& & 0 & & 0  \\
\\[0.01mm]
$R$ &       0.76(3)$^\ast$ && 0.65(2) & & 0.62(4) & &     0.53(8) & & & & &   \\
\\[0.01mm]	
$M$&   0.54(5) && 0.57(3) & & 0.54(5) & &     0.53(8)   & & 0.49(2) & & 0.54(2) & & 0.54(4) 	\\
\\[0.01mm]
$R_F$  &      0.091   && 0.081   & & 0.090 & &        0.14      & & 0.14 & &	0.16 & &	0.24  \\
\\[0.01mm]
$\chi^2$  &   2.77    && 1.78    &&  2.77 &  &         4.78      & & 4.86 & &	3.58 & &	8.74  \\
\\[0.01mm]
 \hline
 \hline
\end{tabular}
\label{table:III}
\end{table*}

\begin{table}[htbp]
\caption {Comparison of the observed magnetic structure factors, $F^2_{obs}$, of magnetic satellite reflections ({\itshape h, k, l})$_M$ and the structure factor $F^2_{calc}$ calculated magnetic structure factors of the spiral model 3 (see Table III).}
\centering
\begin{tabular} {c c c c c c c c}
\hline\hline
\\[0.05mm]
 ($h$ & $k$ & $l$)$_M$ &  &  $F^2_{obs}$ $\pm \sigma$ \hspace{5 mm} & & $F^2_{calc}$ \hspace{5 mm}& \\[0.5ex]
 \hline
  0.5  & 0.228 &  0 & \hspace{5 mm} & 0.029 $\pm$ 0.005 & \hspace{5 mm} & 0.029  &	\\

 $\overline{0.5}$ &   0.228   &  0 & \hspace{5 mm}&  0.026 $\pm$  0.007 &  & 0.029 &	 \\

 $\overline{0.5}$  &  0.228   & 1 &   & 0.168 $\pm$  0.020 &  & 0.141	 &	       \\

 $\overline{0.5}$   &   0.228   & $\overline{1}$ &   &0.190 $\pm$  0.023 &  & 0.141	&	 \\

  0.5 &  0.228 & $\overline{1}$  & & 0.184 $\pm$  0.021	&  &       0.141 &	\\

  0.5  & 0.228  & 1   & &0.170 $\pm$ 0.024 &  &  0.141	 &\\

  0.5 &  0.228  &$\overline{2}$  & & 0.077 $\pm$ 0.015	&  & 0.073 &	\\

 $\overline{0.5}$  & 0.228  & 2  &  & 0.063 $\pm$ 0.012	 &  & 0.073 & \\

  1.5 &  0.228  & 0  &  & 0.049 $\pm$ 0.013 	 &  & 0.075 &	\\

 $\overline{1.5}$ &  0.228 &  0  & & 0.039  $\pm$ 0.008	&  & 0.075 &	\\

  0.5  & 0.228 & $\overline{3}$  & & 0.039 $\pm$ 0.023 &  &	0.075 &	\\

$\overline{0.5}$ &  0.228 &  3  & & 0.039  $\pm$ 0.020 &  &	0.075 &	\\

$\overline{1.5}$  & 0.228  & 1  & & 0.051  $\pm$  0.025	&  & 0.041 &	\\

  1.5  & 0.228  & 2  & & 0.130 $\pm$ 0.033 &  & 0.125 &   \\

 $\overline{0.5}$ & $\overline{0.773}$ & $\overline{2}$   & &  0.078 $\pm$ 0.029	&  & 0.081 &	\\

  0.5  & $\overline{0.773}$  & 2   & & 0.094 $\pm$ 0.029  &  & 0.081 &\\

 $\overline{1.5}$ &  0.773  & 1   & &  0.113 $\pm$  0.021 &  &  0.089 &\\

  1.5 &  0.228 &  4  &  & 0.087 $\pm$  0.016 &  &	0.086  & \\

 $\overline{1.5}$ &  0.228 & $\overline{4}$  & & 0.088 $\pm$ 0.011 &  & 0.086 &	 \\

 $\overline{0.5}$  & 0.228  & 5 &  & 0.057 $\pm$  0.009 &  &  0.074 &\\
 \hline
 \hline
\end{tabular}
\label{table:IV}
\end{table}

The observed magnetic structure factors were then compared to those generated by the complete set of four magnetic models compatible with the lattice symmetry, which can be specified by the possible permutations of the real and imaginary parts, $R_i$ and $I_i$, of the Fourier coefficients of the magnetic moment along the crystal axes $i$. They include a sinusoidal modulation with only real components $R_xR_yR_z$ (model 1), and three elliptical helices with components $I_xR_yR_z$, $R_xR_yI_z$, and $R_xI_yR_z$  (models 2--4). Note that replacing $R \leftrightarrow I$ yields physically equivalent descriptions. In all cases, the representation analysis fully constrains the spin sequences of the atoms inside the primitive unit cell. The model proposed by Kobayashi and coworkers for LiCu$_2$O$_2$, \cite{kobayashi} in which the phase relation of these moments was chosen arbitrarily, is not compatible with the lattice symmetry and was not considered here. The description of more complex structures such as conical helices would require more than one magnetic representation. Such models are therefore also incompatible with our analysis, which yields a single representation.

The program Fullprof \cite{carvajal} was used to perform least-squares refinements of the three Fourier components of the magnetization in the four symmetry-compatible structures. The outcome is shown in Table III, along with the average moment and the moment amplitude derived from these quantities and the quality-of-fit parameters $R_F$ and $\chi^2$. Low residuals $R_F$ were obtained for models 1--3. For the sine-wave modulated structure (model 1 in Table III) the components along the $b$- and $c$-axes were found to be very similar, while the component along $a$ is slightly reduced. For the helical structures (models 2 and 3 in Table III), we obtained nearly the same values of the three magnetization components as calculated for the sinusoidal structure. In model 2, where the real components are in the $bc$-plane and the imaginary component points along the $a$-axis, the copper moments are rotating in two different planes, which subtend angles of $\pm 45(4)^\circ$ with the $b$-axis and encompass the moments of Cu1 and Cu4, and Cu2 and Cu3, respectively. In model 3, where the real components are in the $ab$-plane and the imaginary component points along the $c$-axis, we also obtained two different rotation planes for the moments of Cu1 and Cu3, and Cu2 and Cu4, respectively, which were found to subtend angles of $\pm 50(4)^\circ$ with the $a$-axis. In Table IV, the magnetic structure factors calculated for the latter model (which is the most likely ground state of NaCu$_2$O$_2$, see Section III) are compared to the experimentally observed ones. Model 4 yields a substantially larger $R_F$ and can hence be excluded.

Figure 3 provides pictorial representations of models 2 and 3. Both structures are elliptical helices by symmetry, but since the real and imaginary Fourier coefficients are identical within the standard deviation of the refinement, the ellipticity is small. For comparison, we also investigated simple circular helices with polarization planes coincident with high-symmetry planes of the crystal lattice (models 5--7 in Table III), but these models yielded unsatisfactory residuals.

\begin{figure}[htbp]
\begin{center}
\includegraphics[width=8cm,angle=0]{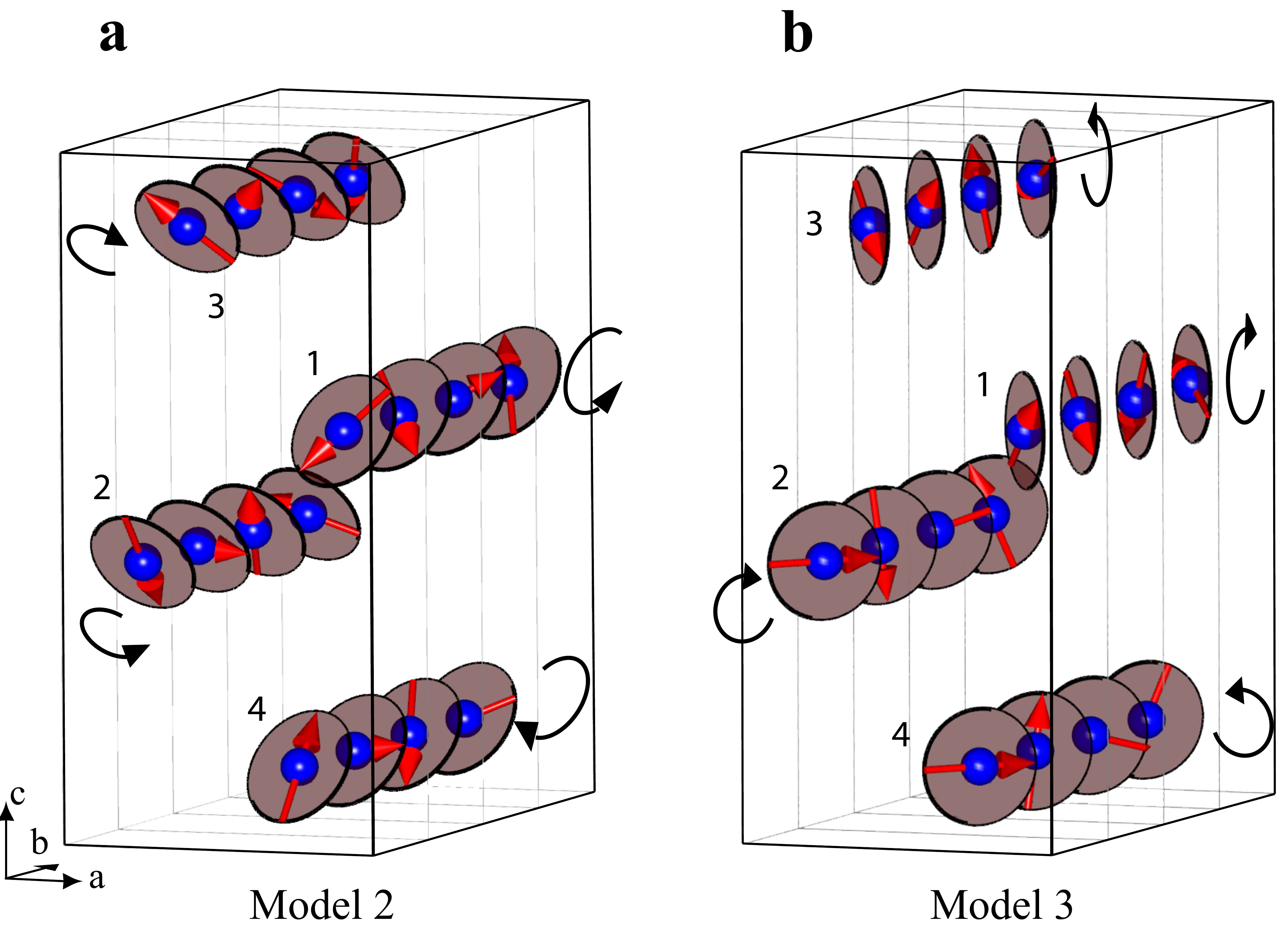}\\
\caption{(Color online) Proposed magnetic structures for NaCu$_{2}$O$_{2}$. Model 3 yields competitive residuals in the refinement of the neutron diffraction data (Table III) and is consistent with other experimental data on NaCu$_2$O$_2$ (see the text). The arrows indicate the sense of rotation of helices on different copper-oxide chains.}
\end{center}
\end{figure}

\section{Discussion and Conclusions}

We first discuss possible limitations of the representation analysis that underlies the choice of magnetic structures we have selected for comparison with the data. The analysis is based on the assumption that the spin Hamiltonian includes terms up to bilinear order in the spin operators. \cite{bertaut} While this is sufficient in the vast majority of magnetic insulators, we note that strong charge and/or orbital fluctuations may lead to higher-order terms that require a modified analysis.\cite{bohnenbuck,reehuis} In view of the large Mott-Hubbard gap of copper-oxide chain compounds \cite{pisarev,malek,matiks} and the large crystal-field splitting of the Cu $d$-orbitals, such fluctuations are expected to be negligible in NaCu$_2$O$_2$. Since recent research has focused attention on ring-exchange terms in the spin Hamiltonian of cuprate spin-ladder compounds, \cite{mikeska,goessling} one might also consider three-spin interactions of moments on directly adjacent copper oxide chains in NaCu$_2$O$_2$. While no quantitative information on the magnitude of such interactions is available, they are expected to be substantially smaller than the bilinear interactions and therefore have no major influence on the magnetic structure. An analysis of the current set of neutron diffraction data based on bilinear exchange interactions therefore appears to be adequate. In any case, an analysis based on a choice of models without recourse to symmetry considerations \cite{kobayashi} seems inadequate.

Three of the four magnetic structures revealed by the representation analysis yield good agreement with the neutron data, and none of them can be singled out based on the diffraction study alone. We therefore discuss these structures in the light of data collected by other experimental probes. Amplitude-modulated states such as the sinusoidal structure of model 1 have been observed in compounds with doped edge-sharing copper oxide chains, which support low-energy charge fluctuations, \cite{raichle} but as mentioned above, such fluctuations are strongly suppressed in Mott insulators such as NaCu$_2$O$_2$. Collinear incommensurate structures with weaker amplitude modulations, such as the soliton lattice observed in spin-Peierls systems in high fields, \cite{kiryukhin} differ from model 1 by the content of higher harmonics, which could not be determined in the present study because of the weakness of the corresponding higher-order Bragg reflections. However, the strongly anisotropic behavior of the Na NMR lineshape for {\bf H} applied along the different crystallographic directions \cite{gippius} is difficult to reconcile with a collinear structure and indicates helical order.

The temperature dependence of the uniform magnetic susceptibility, $\chi$, yields further insight into the magnetic order. Recent measurements on crystals from the same batch as the ones investigated here \cite{leininger} indicate a suppression of $\chi$ for ${\bf H} \parallel b$ and $c$, but not $a$, upon cooling below $T_N$.  Helical states such as model 2, in which the $a$-axis is in the plane of polarization, are inconsistent with these data. In model 3, on the other hand, the $a$-axis subtends a larger angle with the polarization plane than both $b$- and $c$-axes, in qualitative agreement with the low-field susceptibility data. The low symmetry of the polarization plane in this model also explains the apparent absence of spin-flop transitions for fields up to 7 T. \cite{gippius,leininger} Rather than a sharp spin-flop, an external magnetic field along $a$ is expected to induce a gradual rotation of the polarization plane towards the $bc$-plane, which explains the good agreement of high-field NMR data with a model based on $bc$-polarized spirals. \cite{gippius}

The elliptical helix structure with alternating polarization planes shown in Fig. 3b is more complex than the simple circular, $bc$-polarized helix previously identified based on less complete powder neutron diffraction \cite{capogna} and NMR \cite{gippius} data. We stress, however, that these features are mandated by the lattice symmetry and the propagation vector, which are accurately known for NaCu$_2$O$_2$. We also note the alternating sense of rotation of helices propagating along different copper-oxide chains (arrows in Fig. 3). According to Eq. 1, this implies an anti-ferroelectric state in which every chain generates a ferroelectric moment, but the macroscopic electric polarization vanishes. This explains the absence of ferroelectricity in NaCu$_2$O$_2$. \cite{leininger} A small reduction of the dielectric constant below $T_N$ may be indicative of an anti-ferroelectric state. \cite{leininger}

Our results also cast light on the origin of the ferroelectric polarization in LiCu$_2$O$_2$, which exhibits the same lattice symmetry and an incommensurate helix propagation vector of the same form, ${\bf k} =  (0.5, \xi, 0)$, as in NaCu$_2$O$_2$. \cite{masuda} Our representation analysis therefore also applies to LiCu$_2$O$_2$. Since according to Eq. 1 none of the four magnetic states revealed by this analysis supports a macroscopic ferroelectric polarization, we conclude that the ferroelectricity observed \cite{park,yasui1} in LiCu$_2$O$_2$ cannot be of intrinsic origin, and that defects generated by Li--Cu intersubstitution must play a central role. In this respect, our conclusion agrees with those of Ref. \onlinecite{moskvin_prb}, but disagrees with those of Ref. \onlinecite{yasui1}. We emphasize, however, that our findings do not invalidate models according to which helical magnetism generates ferroelectricity. \cite{katsura,sergienko,mostovoy} In the framework of this scenario, substitutional defects may locally lift the compensation of ferroelectric moments in different copper-oxide chains, and a description of the resulting magneto-electric defect pattern may be the key to an explanation of the puzzling magnetic field dependence of the ferroelectric polarization in LiCu$_2$O$_2$. \cite{park,yasui1} An assessment of the possible influence of defects on the ferroelectric properties of LiCuVO$_{4}$, \cite{naito,yasui2} as well as other multiferroics with helicoidal order, \cite{wang} is an interesting subject of further investigation.

\section*{ACKNOWLEDGMENTS}
We acknowledge useful discussions with Ph. Leininger, G. McIntyre, and J. Rodriguez Carvajal. We also acknowledge financial support by the DFG under grant SFB/TRR 80.

\end{document}